\documentclass[twocolumn,english,prd,nofootinbib,superscriptaddress]{revtex4-1}
\usepackage[T1]{fontenc}
\usepackage[latin9]{inputenc}
\usepackage{babel}
\usepackage{bm}
\usepackage{amsmath}
\usepackage{graphicx}
\usepackage{wasysym}
\usepackage[unicode=true,pdfusetitle,
 bookmarks=true,bookmarksnumbered=false,bookmarksopen=false,
 breaklinks=false,pdfborder={0 0 1},backref=false,colorlinks=false]
 {hyperref}

\usepackage{caption}
\newcommand{\bmf}[1]{{\bm{#1}}}

\makeatletter

\providecommand{\tabularnewline}{\\}
\newcommand{\lyxdot}{.}

\usepackage{xcolor}

\@ifundefined{showcaptionsetup}{}{%
 \PassOptionsToPackage{caption=false}{subfig}}
\usepackage{subfig}
\makeatother

\begin{document}

\title{Magnetic field production at a first-order electroweak phase transition}

\author{Yiyang Zhang} 
\affiliation{Department of Physics, McDonnell Center for the Space Sciences,
Washington University, St. Louis, Missouri 63130, USA}
\author{Tanmay Vachaspati}
\affiliation{Physics Department, Arizona State University, Tempe, Arizona 85287, USA}
\author{Francesc Ferrer}
\affiliation{Department of Physics, McDonnell Center for the Space Sciences,
Washington University, St. Louis, Missouri 63130, USA}

\begin{abstract}
	We study the generation of magnetic field seeds during a
first-order electroweak phase transition, by numerically evolving the classical 
equations of motion of the bosonic electroweak theory on the lattice. 
The onset of the transition is implemented by the random 
nucleation of bubbles with an arbitrarily oriented Higgs field
in the broken phase.
We find that about $10$\% of the latent heat is converted into magnetic
energy, with most of the magnetic fields being generated in the last stage 
of the phase transition when the Higgs oscillates around the 
true vacuum.
The energy spectrum of the magnetic field has a peak that
shifts towards larger length scales as the phase transition unfolds.
By the end of our runs the peak wavelength is 
of the order of the bubble percolation scale, or about a third of our 
lattice size. 
\end{abstract}

\maketitle

\section{Introduction\label{sec:Introduction}}

Magnetic fields are pervasive in the Universe. Micro-gauss fields coherent on
scales up to ten kpc have been detected in nearby spiral galaxies, such as 
the Milky Way and in higher redshift galaxies (e.g. \cite{Kronberg:1993vk}).
Similar stochastic magnetic fields are found in clusters of galaxies. Such fields
are believed to result from the dynamo amplification of weak magnetic field
seeds, whose origin remains a mystery (see 
e.g.~\cite{Kandus:2010nw,Durrer:2013pga,Subramanian:2015lua}). 
Recent observational evidence that even the
intergalactic medium in voids is pervaded by a weak (below femto-gauss 
strength) magnetic 
field~\cite{Neronov:1900zz,Tavecchio:2010mk,Tashiro:2013ita,Chen:2014qva,Chen:2014rsa,Biteau:2018tmv}, 
points to a primordial origin of the field seed, 
since it is difficult to account for a field that fills the volume in the 
void regions by astrophysical processes in the late 
universe~\cite{Furlanetto:2001gx,Bertone:2006mr} (but
see~\cite{Durrive:2015cja, Beck:2012cs}). 
Furthermore, the study of the diffuse gamma-ray background provides clues
for the existence of a nonzero helical component in the 
intergalactic magnetic field~\cite{Tashiro:2013ita,Chen:2014qva,Duplessis:2017rde}. Interestingly,
while most magnetic field generation mechanisms discussed so far produce 
nonhelical fields, physical processes associated with electroweak 
baryogenesis imply fields that are 
helical~\cite{Cornwall:1997ms,Vachaspati:2001nb}. 
This is important, since conservation of magnetic helicity plays an
important role in the evolution of primordial fields and leads, via inverse
cascade, to a larger coherence scale than for nonhelical 
fields~\cite{Christensson:2000sp}.

The production of helical fields is related to the violation of baryon
plus lepton (B+L) number during the electroweak phase transition (EWPT). 
To change B+L requires a change in the Chern-Simons number of the 
electroweak gauge fields, which proceeds through ``sphaleron'' configurations.
The decay of the sphaleron releases helical magnetic fields, and this is 
borne out by both numerical simulations~\cite{Copi:2008he} and analytical 
arguments~\cite{Chu:2011tx}. 
Generically, helical magnetic fields are produced
at the electroweak phase transition if the 
Chern-Simons number changes during the phase transition~\cite{Zhang:2017plw}.
Inhomogeneities in the Higgs field can also give rise to the generation of
helical magnetic fields as shown in~\cite{DiazGil:2007dy,DiazGil:2008tf} in
the context of low-scale electroweak hybrid 
inflation~\cite{GarciaBellido:2002aj}.

The study of the properties of cosmological magnetic fields can, thus, open
a window to the early universe and very high energy particle interactions. 
To investigate the epoch of magnetogenesis and identify the mechanism 
at the origin of the primordial seeds we need to make full use of the
statistical properties of the magnetic fields matching theoretical predictions
with observations~\cite{Caprini:2015gga}. These are encoded in the 
power spectrum which, for Gaussian random fields, entirely describes the 
statistical properties of the fields. In this respect, 
the helicity spectrum can be recovered from observations of the 
diffuse gamma-ray background~\cite{Chen:2014qva}.

On the theoretical front, it has been shown that a measurement of the sign of 
the helicity can 
distinguish fields produced during electroweak baryogenesis from those 
generated at an earlier epoch associated with leptogenesis~\cite{Long:2013tha}.
However, a detailed study of the statistical properties of the magnetic fields is required
to narrow the gap between theoretical studies of the microscopic properties 
and the macroscopic properties that can be experimentally observed (field 
strength and coherence scale), as was done for the case of low-scale hybrid
inflation in~\cite{DiazGil:2007dy,DiazGil:2008tf}.

In this paper, we study the dynamics of a first-order EWPT by numerically
evolving the classical bosonic electroweak theory, and we examine the
properties of the generated magnetic field. In the Standard Model (SM), the
EWPT is first order only if the mass of the Higgs boson lies below
$m_h \lesssim 70$ GeV~\cite{Bochkarev:1987wf,Kajantie:1995kf}.  
Since the experimentally observed mass violates this 
bound~\cite{ATLAS:2012ae,Chatrchyan:2012tx}, we work under the
assumption that there is physics beyond the SM that influences the character
of the phase transition. Although the details of the particular SM extension
are not important for our purpose, we have in mind models that make 
electroweak baryogenesis viable~(see 
e.g.~\cite{Riotto:1999yt,Quiros:1999jp,Dine:2003ax,Morrissey:2012db} 
for reviews). In that case, the properties of the cosmological magnetic
fields can be related to the observed baryon asymmetry of the 
universe~\cite{Vachaspati:2001nb}.
Furthermore, a cosmological first-order phase transition can also be a 
source of gravitational waves (GWs)~\cite{Witten:1984rs,Hogan:1986qda}.
In this respect, 
large-scale numerical simulations of a scalar field theory on the lattice
have recently become available to verify the production of GW radiation 
during a first-order phase transition~\cite{Cutting:2018tjt, Ellis:2019tjf, Kharzeev:2019rsy}.
Our results bring another handle to probe the universe at the time of the
EWPT: the observation of cosmological magnetic fields. 

When the phase transition occurs, the Higgs field will leave the
symmetric phase and gradually settle down around the true vacuum,
$\vert\Phi\vert=\eta$. In order to mimic
this behavior on the lattice, we introduce a phenomenological damping term
for the Higgs boson. The modified electroweak evolution equations preserve
gauge invariance and satisfy the Gauss constraints. They 
are reviewed in Sec.~\ref{sec:Theory}, where we also outline
the calculation of the magnetic field spectrum from our lattice simulations.

Magnetic fields are produced as a result of nonvanishing gradients 
of the Higgs field~\cite{Vachaspati:2001nb}.
In a first-order phase transition, bubbles
are randomly nucleated as a result of quantum tunneling, and subsequently
expand and collide with each other, producing an out-of-equilibrium
environment. 
Magnetic fields, will be produced when bubbles collide. 
We argue that
the details of the short period of bubble nucleation are not of major concern,
and we can safely mimic the quantum tunneling by a simple random procedure,
controlled by a parameter $p_{B}$, the nucleation probability. 
The specifics of our numerical implementation are presented in 
Sec.~\ref{sec:Numerical-Simulation}.

To gain intuition about the general features of the magnetic fields generated
from bubble collisions, we first consider in Sec.~\ref{sec:nrand-dist} a set of
constrained simulations with a regular array of bubbles so that we can control
the size of the colliding bubbles. 
In Sec.~\ref{sec:rand-nucl} we allow for bubbles to nucleate at random locations
in the unbroken phase and we present the 
spectrum of the magnetic field induced during a first-order EWPT. 
At the end of the bubble collision stage, we find that the field strength 
is about $10^{23}$ gauss, with a comoving correlation length of about 2 cm. If the fields are nonhelical, the field strength will evolve to 
$\sim 10^{-15}$ gauss at present~\cite{Banerjee:2004df}.
Finally, we discuss and summarize our results in Sec.~\ref{sec:conclusion}.

\section{Theoretical Formulation\label{sec:Theory}}

\subsection{Classical equations and Higgs damping\label{subsec:Classical-eqns}}

We consider the classical bosonic electroweak theory, which includes
the Higgs doublet $\Phi$, the SU(2)-valued gauge fields $W_{\mu}^{a}$
and the U(1) hypercharge field $B_{\mu}$. The Lagrangian is given
by 
\begin{equation}
\mathcal{L}=-\frac{1}{4}W_{\mu\nu}^{a}W^{a\,\mu\nu}-\frac{1}{4}B_{\mu\nu}B^{\mu\nu}+\vert D_{\mu}\Phi\vert^{2}-\lambda(\vert\Phi\vert^{2}-\eta^{2})^2,\label{eq:lagrangian}
\end{equation}
where 
\begin{equation}
D_{\mu}=\partial_{\mu}-\frac{i}{2}g\sigma^{a}W_{\mu}^{a}-\frac{i}{2}g'B_{\mu}.\label{eq:cov-deriv}
\end{equation}

Since we will evolve the field equations using the standard Wilsonian approach for lattice gauge 
fields~\cite{Ambjorn:1990pu,Moore:1996wn,Tranberg:2003gi,GarciaBellido:2003wd},
it is advantageous to use the temporal gauge, $W_0^a = B_0=0$, which allows a
simple identification of the canonical momentum. Our implementation of the 
lattice equations can be found in~\cite{Rajantie:2000nj,Zhang:2017plw}, with
the addition of a Higgs damping term that we now discuss. 
The classical equations of motion (EOMs) in the continuum are given by:
\begin{eqnarray}
\partial_{0}^{2}\Phi = && D_{i}D_{i}\Phi-2\lambda(\vert\Phi\vert^{2}-\eta^{2})\Phi-\gamma\Phi\, \partial_{0}\ln\vert\Phi\vert,
\label{eq:eom-Higgs}\\
\partial_{0}^{2}B_{i} = && -\partial_{j}B_{ij}+g'\, \text{Im} [\Phi^{\dagger}D_{i}\Phi ],
\label{eq:eom-u1}\\
\partial_{0}^{2}W_{i}^{a} = && -\partial_{k}W_{ik}^{a}-g\, \epsilon^{abc}W_{k}^{b}W_{ik}^{c}
 +g\, \text{Im}[\Phi^{\dagger}\sigma^{a}D_{i}\Phi ].
\label{eq:eom-su2}
\end{eqnarray}
The solutions are subject to the Gauss constraints: 
\begin{eqnarray}
	\partial_{0}\partial_{j}B_{j} && - g'\, \text{Im}[\Phi^{\dagger}\partial_{0}\Phi ] = 0,
\label{eq:gauss-u1}\\
	\partial_{0}\partial_{j}W_{j}^{a} &&+g\, \epsilon^{abc}W_{j}^{b}\partial_{0}W_{j}^{c} - 
g\, \text{Im}[\Phi^{\dagger}\sigma^{a}\partial_{0}\Phi ] = 0.
\label{eq:gauss-su2}
\end{eqnarray}

The term proportional to $\gamma$ in Eq.~(\ref{eq:eom-Higgs}) gives rise to a linear damping 
of the magnitude of the Higgs field.
Its purpose is to attenuate the radial oscillations of the
Higgs field within a reasonable simulation time. The term also simulates
the effects of energy losses due to Higgs decays into 
fermion-antifermion pairs. In principle, one could include separate damping 
terms for each component of the Higgs field (e.g. to account for the 
decays of gauge bosons).
However, this is harder to implement since we want to respect gauge invariance and the conservation of electric charge. 

The form of the damping term in Eq.~(\ref{eq:eom-Higgs}) is motivated 
by considering the analogy of a 2D simple harmonic oscillator
with linear damping in the radial direction. In polar coordinates, the EOMs are
\begin{align}
\ddot{r}-r\dot{\theta}^{2}+c\dot{r}+r & =0,\\
r^{2} \dot{\theta} & =L,
\end{align}
where, $r=\vert\bmf{r}\vert$, $\bmf{r}=\left(x,y\right)$. The second
equation is the conservation of angular momentum and corresponds to
the conservation of charge in the field theory.
If the two equations are transformed into
Cartesian coordinates, we find
\begin{equation}
\ddot{\bmf{r}}+c\bmf{r} \frac{d}{dt}\ln r+\bmf{r}=0,\label{eq:sho-damping-xy-comp}
\end{equation}
which suggests the modification of Eq. (\ref{eq:eom-Higgs}).

To check that the damping term is not in conflict with the Gauss constraints, let us
write a general additional term for the Higgs EOM, 
\begin{align}
\partial_{0}^{2}\Phi_{a} & =D_{i}D_{i}\Phi_{a}-2\lambda(\vert\Phi\vert^{2}-\eta^{2})\Phi_{a}+\Xi_{a},\nonumber \\
 & a=1,2,3,4,\label{eq:general-modified-Higgs-EOM}
\end{align}
where $\Xi=\left(\Xi_{1}+i\Xi_{2},\Xi_{3}+i\Xi4\right)^{T}$ is a doublet that causes damping.
The Gauss constraints should be satisfied throughout the evolution assuming
that they are initially satisfied, which requires that 
the time derivatives of Eqs.~(\ref{eq:gauss-u1}) and~(\ref{eq:gauss-su2}) 
should vanish.  
Differentiating the two equations, and using the modified EOM 
Eq.~(\ref{eq:general-modified-Higgs-EOM}) for the Higgs field, together with 
Eqs.~(\ref{eq:eom-u1}) and~(\ref{eq:eom-su2}), to replace the second-order
time derivatives, we obtain
\begin{eqnarray}
\text{Im}[\Phi^{\dagger}\Xi] & = & 0,\\
\text{Im}[\Phi^{\dagger}\sigma^{i}\Xi] & = & 0.
\end{eqnarray}
The solution to these equations, which ensures the Gauss constraints, is 
\begin{equation}
\Xi\propto \Phi,
\label{eq:Higgs-add-term-sol}
\end{equation}
and this gives the damping term in Eq.~(\ref{eq:eom-Higgs}).

To simulate a first-order electroweak phase transition we also 
need to 
include the quantum nucleation of broken phase bubbles, which we now discuss. 

\subsection{Bubble profile}
\label{subsec:nucleation}

As mentioned above, the SM does not admit a first-order 
electroweak phase transition. 
However, we expect that for our purposes the Higgs potential as given in 
Eq.~(\ref{eq:lagrangian}) captures the dynamics of the phase
transition in an extension of the SM that allows for a viable 
electroweak baryogenesis.  
The equations of motion, with the initial condition $\Phi=\dot{\Phi}=0$,
are then supplemented by bubble nucleation events that occur randomly in 
regions where the symmetry is unbroken. 

In numerical simulations, we use a simple method for randomly nucleating bubbles of the broken
phase to mimic the tunneling effect. The sites of the lattice can be 
numbered as a sequence $S$ from 1 to $N^3$.
At each time step of the simulation, we first randomly shuffle this sequence, 
and then we select sites from this shuffled sequence $S_{\text{rand}}$,
with probability $p_B$ that controls the rate of nucleation on the lattice. 
The selected sites are stored $S_{\text{select}}$, which contains
$\sim p_BN^3$ elements. We scan each site 
$s_i$ in $S_{\text{select}}$ and
if the Higgs field in all lattice sites within a radius $r_0$ from $s_i$ 
is still in the symmetric phase,
then a new bubble centered at $s_i$ is nucleated at this time step,
otherwise, the site $s_i$ is skipped. 
The random shuffle procedure of all sites 
guarantees that the nucleation procedure is unbiased in any direction.

At every nucleation event we set up a bubble with a spherically symmetric
profile $| \Phi | = f(r)$, determined  
by demanding that the nucleation process conserves energy. 
This requires that the energy change due to bubble nucleation should vanish. 
Therefore,
\begin{equation}
0 = 4\pi \int_0^\infty r^2 dr \left[ ( \partial_r f )^2 - 2 \lambda \eta^2 f^2 + \lambda f^4  \right ].
\end{equation} 
This equation can be satisfied by choosing
\begin{equation}
\partial_r f = - \sqrt{\lambda} ( 2 \eta^2 - f^2 )^{1/2} f.
\end{equation}
Rescaling $\rho = \sqrt{2\lambda}\, \eta r$ and $F = f/(\sqrt{2}\, \eta)$ gives
\begin{equation}
\partial_\rho F = -( 1 - F^2 )^{1/2} F
\end{equation}
and the solution is
\begin{equation}
F(\rho ) = \frac{2C e^{-\rho }}{1+C^2 e^{-2\rho}}.
\end{equation}
The integration constant, $C$, is fixed by requiring that the center of
the bubble be in the true vacuum, so $f(0) = \eta$ ($F(0)=1/\sqrt{2}$). This leads
to $C = \sqrt{2}\pm 1$, and we choose the smaller value so as to have a gentler
bubble profile. 
The final solution for the bubble profile is
\begin{equation}
f(r) = \eta \, \frac{[1+ (\sqrt{2}-1)^2]\, e^{-m_H r/\sqrt{2}}}
                       {1+ (\sqrt{2}-1)^2 e^{-\sqrt{2}\, m_H r}}
                       \label{eq:exp-profile}
\end{equation}
where $m_H = 2 \sqrt{\lambda} \, \eta$ is the Higgs mass.
\footnote{One issue with this bubble profile function is that it has a kink at $r=0$. 
For example, along the $x$ axis and close to the origin it behaves as 
$\exp(-|x|)$. This is not a problem numerically as finite differences will 
not resolve the kink.}

In this way we have fixed the bubble profile using the (stronger) requirement that 
bubble nucleation conserve energy {\it locally}. The direction of the 
Higgs within the
bubble is assumed to be uniform. Since the vacuum manifold 
-- zeros of the Higgs potential -- defines a three sphere, the direction of 
the Higgs is chosen by randomly selecting a point with a uniform distribution 
on the three sphere. Different bubbles will have different Higgs field orientations.

\subsection{Definition of electromagnetic field}
\label{emdefn}

Once the Higgs field leaves the symmetric phase ($\Phi=0$), we can define the 
electromagnetic field as follows: 
\begin{equation}
A_\mu = \sin \theta_w n^a W^a_\mu +  \cos \theta_w B_\mu,
\label{eq:amu}
\end{equation}
where,
\begin{equation}
n^a \equiv -\frac{\Phi^\dagger \sigma^a \Phi}{\eta^2}
\end{equation}
indicates the direction of the Higgs field, and $\theta_w$ is the weak mixing angle.
The corresponding field strength is constructed 
as~\cite{tHooft:1974kcl,Vachaspati:1991nm}
\begin{align}
\nonumber
A_{\mu\nu} &= \sin \theta_w n^a W^a_{\mu\nu} + \cos \theta_w B_{\mu\nu} \\ 
&- i\frac{2}{g \eta^2} \sin \theta_w \left[ (D_\mu\Phi)^\dagger (D_\nu \Phi) - (D_\nu\Phi)^\dagger (D_\mu \Phi) \right].
\label{eq:amunu}
\end{align}

\subsection{Magnetic energy spectrum}
\label{subsec:Magnetic-energy-spectrum}

We assume that the magnetic field after production is a statistically 
homogeneous and isotropic, Gaussian-distributed vector field. The field
can then be described in terms of the equal time correlation function, 
which we write, following the conventions in~\cite{Brandenburg:2017neh, Brandenburg:2018ptt}, 
as 
\begin{equation}
\langle B_i^* (\bmf{k},t )B_j(\bmf{k}',t)\rangle 
= (2\pi )^3 \delta^{(3)}(\bmf{k}-\bmf{k}' )F_{ij} (\bmf{k},t),
\label{eq:spectrum-function}
\end{equation}
where $ B_i (\bmf{k},t )$ is the Fourier transform of $ B_i (\bmf{x},t )$
with the convention
\begin{eqnarray}
B_{i}\left(\bmf{k},t\right) & =&\int d^{3}xB_{i}\left(\bmf{x},t\right)e^{+i\bmf{k}\cdot\bmf{x}},\\
B_{i}\left(\bmf{x},t\right) & =&\int \frac{d^{3}k}{\left(2\pi\right)^{3}} B_{i}\left(\bmf{k},t\right)e^{-i\bmf{k}\cdot\bmf{x}}.
\end{eqnarray}
The spectrum, $ F_{ij} (\bmf{k},t)$, can be divided into a symmetric 
(nonhelical) part and an antisymmetric (helical) part, 
\begin{equation}
\frac{F_{ij} (\bmf{k},t)}{(2\pi )^3}
= (\delta_{ij}-\hat{k}_{i}\hat{k}_{j}) \frac{E_{M}(k,t)}{4\pi k^{2}}+i\epsilon_{ijl}k_{l}\frac{H_{M}(k,t)}{8\pi k^{2}}.
\label{eq:spectrum-decomp}
\end{equation}

The mean magnetic energy density can be written as 
\begin{equation}
\rho_B \left(t\right) =  \frac{1}{2} \langle {\bf B}^2({\bf x},t) \rangle
=\int_{0}^{\infty}E_{M}\left(k,t\right)dk.
\label{eq:energy-density}
\end{equation}

The average wave number 
\begin{equation}
k_{\text{mean}}\left(t\right)=\frac{\int_{0}^{\infty}kE_{M}\left(k,t\right)dk}{\int_{0}^{\infty}E_{M}\left(k,t\right)dk},
\label{eq:def-kmean}
\end{equation}
provides a characteristic of the energy distribution in a given field
configuration.

We implement discretized versions of the expressions above on a 
three-dimensional lattice containing $N$ nodes separated a distance $\Delta x$
along each spatial dimension of length $L= N \Delta x$. Every lattice
point is labeled by a triplet of integers, $\bmf{X}$, each ranging from
$0$ to $N-1$: $X_{i}\in\left\{ 0,1,...,N-1\right\} $ , for $i=1,2,3$. The
corresponding Fourier space is also described by triplets of integers,
$K_{i}$, of the same form. The
largest wavelength along a particular direction, 
corresponding to the smallest momentum, is of order $L$, while the smallest wavelength that can be effectively 
described is determined by the spacing $\Delta x$. Allowing for modes
traveling in the positive and negative directions, a given physical 
wave number corresponds to a triplet $K'_{i}$:
\begin{equation}
k_i=2\pi K'_i /L,
\label{eq:wave-number}
\end{equation}
where, 
\[  K'_i =  \left\{
\begin{array}{ll}
      K_i & K_i\leq N/2 \\
      K_i-N & K_i > N/2. \\
\end{array} 
\right. \]
For the physical wavelength associated to a momentum with magnitude $k$ 
we have
\begin{displaymath}
\lambda_{k}=2\pi / k=N\Delta x / K',
\end{displaymath}
where the magnitude, $K'=|{\bf K'}|$, ranges from 0 to $\sqrt{3}N/2$ on a 
cubic lattice. We divide this range into bins of size $\Delta K'$,
each with center value $K'_c$. 

Then Eq. (\ref{eq:energy-density}) can be approximated by
\begin{equation}
\rho_B(t) = \sum_{K'_c} E_M(K'_c,t) \Delta K',
\end{equation}
where, 
\begin{equation}
E_M(K'_c,t) \equiv \frac{1}{2\Delta K'} \left ( \frac{1}{L} \right )^6 
\sum_{c^{th}\text{ bin}} B_i^{*}( {\bf K'}, t ) B_{i} ({\bf K'},t).
\label{eq:E_M}
\end{equation}

Also, the discrete Fourier transform is given by
\begin{align}
B_{i}\left(\bmf{K}\right) & = (\Delta x)^3 
\sum_{\bmf{X}=0}^{N-1}B_{i}\left(\bmf{X}\right)\exp\left[+2\pi i\frac{\bmf{K}\cdot\bmf{X}}{N}\right]\label{eq:dft}\\
B_{i}\left(\bmf{X}\right) & = \frac{1}{L^3}  \,
\sum_{\bmf{K}=0}^{N-1}B_{i}\left(\bmf{K}\right)\exp\left[- 2\pi i\frac{\bmf{K}\cdot\bmf{X}}{N}\right]\label{eq:inv-dft}.
\end{align}

\section{Numerical Simulation\label{sec:Numerical-Simulation}}

As mentioned above, we follow the strategy 
in~\cite{Rajantie:2000nj,Zhang:2017plw} to evolve the electroweak EOMs on
the lattice. Our code is based on the 
LATfield2\footnote{http://github.com/daverio/LATfield2} 
library~\cite{David:2015eya}, and the linear algebra operations are performed
with the help of the Eigen\footnote{http://eigen.tuxfamily.org}
library~\cite{eigenweb}.
Our simulations use periodic boundary conditions and the dimensionless
constants entering the EOMs are fixed to their physical values:
$g=0.65$, $\sin^2\theta_w=0.22$, $g'=g\tan\theta_{w}$ and $\lambda=0.129$. 
The spatial and time spacing are chosen to be $\Delta x=0.25$, 
$\Delta t=\Delta x/4=0.0625$, respectively. 
The dimensionful vacuum expectation value of the Higgs, denoted by $\eta$, 
is $174.13~{\rm GeV}$.  
In our numerical code we set $\eta=1$,
so that $\eta \Delta x = 0.25$, and then 
$m_H \Delta x = 2\sqrt{\lambda} \eta \Delta x =0.18$
where $m_H$ is the mass of the Higgs. This choice of lattice spacing gives us
enough resolution to ensure that we capture all the dynamics. 
For instance, since $m_H \Delta x = 0.18$, momenta 
of order $m_H$ are well resolved. The bulk of our simulations is performed on a 
lattice with size $N=256$, although we use a larger lattice for 
several runs in Sec.~\ref{sec:nrand-dist}.
We denote by $T$ the (integer) time step number, 
and the physical time $t$ is $t=T\Delta t$. 

\begin{figure}
\includegraphics[width=0.2\textwidth]{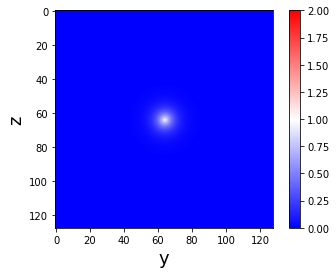}
\includegraphics[width=0.2\textwidth]{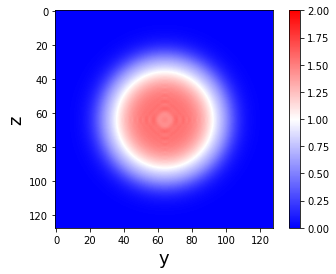}
\caption{
Two-dimensional slice showing the evolution of 
	$\vert \Phi \vert^2/\eta^2$
	for one bubble at time step $T=0$ (left) and $T=140$ (right). 
	Blue-colored regions correspond to $\vert\Phi\vert \ll \eta$, 
	red indicates $\vert\Phi\vert \gg \eta$, and $\vert\Phi\vert \approx \eta$
	in the white regions.}
\label{fig:one-bubble}
\end{figure}

\begin{figure}
\includegraphics[width=0.2\textwidth]{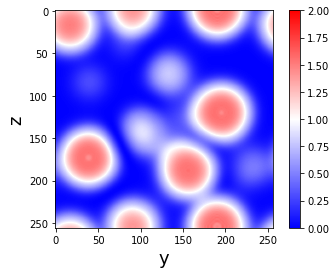}
\includegraphics[width=0.2\textwidth]{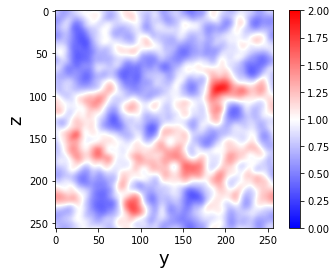}
\caption{ 
Two-dimensional slice showing the evolution of $\vert \Phi \vert^2/\eta^2$
	for randomly nucleated bubbles. The left panel is at time step $T=140$ and 
	the right panel at $T=2000$. 
	Blue colored regions correspond to $\vert\Phi\vert \ll \eta$, 
	red indicates $\vert\Phi\vert \gg \eta$, and $\vert\Phi\vert \approx \eta$
	in the white regions.
}
\label{fig:many-bubbles}
\end{figure}

The bubble profile function, Eq.~(\ref{eq:exp-profile}), does not have 
any free parameters and its tail has infinite extent, which we truncate
on the lattice as follows. 
We define the symmetric phase to correspond to 
locations where 
$|\Phi| \le 0.01\eta$. With this prescription, the ``size'', $r_0$, of the bubble
turns out to be $\eta r_0=9.0$ ($m_H r_0 \approx 6.5$), since the profile in Eq.~(\ref{eq:exp-profile})
falls below $0.01\eta$ for $r > r_0$. With our lattice parameters, this gives 
$r_0$ to be $36 \Delta x$. We use this value to prevent the nucleation
of new bubbles within existing ones: 
a bubble can only be nucleated at a particular site
if all lattice points within a distance $r_0$ are still
in the symmetric phase ($\vert\Phi\vert\le0.01\eta$).  
Once a bubble is nucleated, it
will expand and collide with other bubbles if there are any in the
vicinity. The expansion of a single bubble is shown in  
Fig.~\ref{fig:one-bubble},  
while Fig.~\ref{fig:many-bubbles} shows the evolution and collision
of several randomly generated bubbles.

Two additional inputs required for our runs 
are the Higgs damping $\gamma$, defined in 
Eq. (\ref{eq:eom-Higgs}),
and the nucleation probability $p_{B}$, which determines the probability of bubble
nucleation per lattice site per time step. 
These
two parameters cannot be determined within the model we are considering,
and thus we will compare the results by varying the two parameters.
We consider several values in the range $0 \leq \gamma \leq 0.01$, including
the experimentally measured decay width of Higgs boson, 
$\gamma\sim\Gamma_{\rm Higgs}\sim 4.07 \times 10^{-3}\text{ GeV}$~\cite{Patrignani:2016xqp},
which corresponds to $\gamma \sim 2.34\times 10^{-5}$ in our lattice units. 
$p_B$ is chosen to be in the range $10^{-8} \le p_B \le 10^{-3}$ in our simulations.  

Since we are concerned with the generation of magnetic fields during the 
electroweak phase transition, we need a criterion to determine when the phase
transition is completed.  
Our strategy is to compute the minimum
$\vert\Phi\vert^{2}$ among all the lattice sites at each time step. 
To avoid spurious fluctuations, we work with the ten-step moving average 
of $\vert\Phi\vert_{\text{min}}^{2}$, denoted as
$\vert\Phi\vert_{\text{MA10}}^{2}$, and we stop the simulation at the
first time step $T_{\text{stop}}$ when 
$\vert\Phi\vert_{\text{MA10}}^{2} > 0.25 \eta^{2}$.
In this manner, we ensure that the Higgs field is away from the 
symmetric phase. 

One caveat of our formalism is that 
our field equations do not include the effects of other charged particles
that might be present or generated at the time of the phase transition.

\section{Test runs with nonrandom bubble distributions}
\label{sec:nrand-dist}

A single expanding bubble in our analysis does not generate magnetic fields.
This can be verified from the field equations since
the gauge field currents [right-hand sides of Eqs.~(\ref{eq:eom-u1})
and (\ref{eq:eom-su2})] vanish for a spherically symmetric expanding 
bubble. 
Hence, the simplest setup where we can observe the generation of 
electromagnetic fields involves the collision of two bubbles.

\begin{figure}
\includegraphics[width=0.23\textwidth]{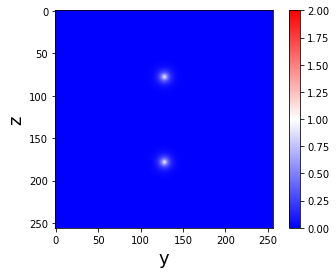}
\includegraphics[width=0.23\textwidth]{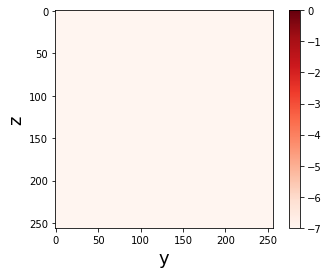}
\includegraphics[width=0.23\textwidth]{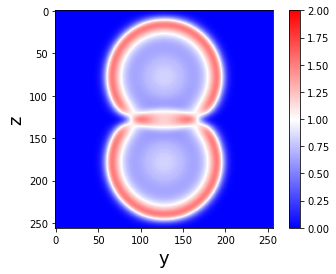}
\includegraphics[width=0.23\textwidth]{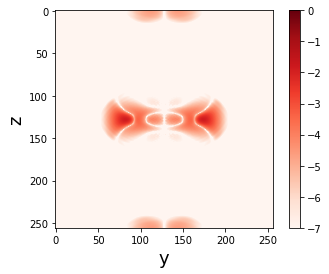}
\caption{
({\it Top}) 2D slice with $x=0$ snapshot of two-bubble collision at $T=0$. The 
top-left plot shows the distribution of $\vert\Phi\vert^{2}/\eta^{2}$ and the top-right
plot shows magnetic energy density on a log-scale.
The two bubbles are still in free expansion, thus no magnetic field is generated. 
({\it Bottom}) Same as the top plots but at $T=240$. A ring-shaped magnetic field 
is generated at the location of the bubble intersection. The magnetic field at the 
upper and lower boundaries appears because of the second collision of the two 
bubbles due to periodic boundary conditions on the lattice. 
The size of the lattice is $N=256$. 
The energy density of the magnetic field is measured in units of $m_H^4$.
}
\label{fig:two-bubble-expand}
\end{figure}

Accordingly, we
nucleate two bubbles along the $z$ axis at $T=0$ and let them expand
and collide. The initial radius $r_{0}$ of each bubble is fixed, 
$r_0 = 36\Delta x$,
but the initial orientations of the Higgs field inside the bubbles are random.
As shown in~Fig.~\ref{fig:two-bubble-expand}, 
the two bubbles initially expand freely before they collide. Once the 
collision occurs, the process of magnetic field generation can start.
Specifically, as we can see from the bottom panel in 
Fig.~\ref{fig:two-bubble-expand},
the magnetic field is generated at the intersection of the two bubbles.
For this two-bubble configuration, a ring-shaped magnetic field will
be produced, at least initially. Let us also emphasize that, since 
we are using periodic boundary conditions, the two bubbles actually collide
\textit{twice} along the $z$ axis during the expansion as can be seen
in Fig.~\ref{fig:two-bubble-expand}. 

\begin{figure}
\includegraphics[width=0.45\textwidth]{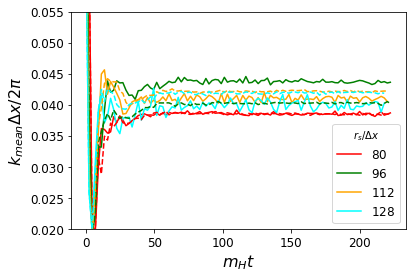}
\caption{\label{fig:nr-adjtime}Evolution of $k_{\text{mean}}\Delta x/2\pi$ for several
nonrandom bubble distributions. The solid lines were computed
	on a lattice of size $N=256$, while for the dashed lines
	$N=400$. 
}
\end{figure}

\begin{figure}
\includegraphics[width=0.45\textwidth]{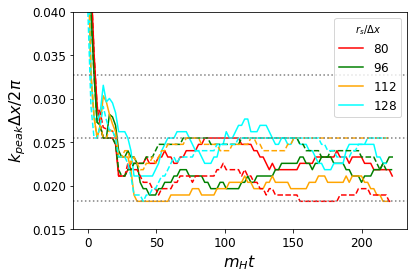}
\caption{\label{fig:nr-adjtime-kpeak}Evolution of $k_{\text{peak}}\Delta x/2\pi$ for several
nonrandom bubble distributions. The data points show the 10-step moving 
average, and the horizontal gray dotted lines correspond to 
the central values of the bins. 
The solid lines were computed
on a lattice of size $N=256$, while for the dashed lines
$N=400$. 
}
\end{figure}

To better understand the general features of the magnetic field resulting from 
multiple bubble collisions, we start with a constrained simulation in
which we control the initial separation of the colliding bubbles. To this effect,
we consider a {\it regular} array of bubbles with centers separated by
a fixed distance $r_s$, which are all simultaneously 
nucleated at $T=0$. No additional bubbles are nucleated at $T>0$. 
In this case, the percolation size $r_p$ of the bubbles can be estimated by $r_p \approx r_s$.
We consider different
values for the initial separation, $r_{s}/\Delta x=80,96,112,128$,
and two lattice sizes, 
$N=256, 400$.
The magnetic field is calculated using 
Eq.~(\ref{eq:amunu}), and then Fourier transformed to obtain the power 
spectrum.
We show the time evolution, in units of $m_H t$, of the mean wave number, 
$k_{\text{mean}} \Delta x/2\pi$,
in Fig.~\ref{fig:nr-adjtime},  
and the peak of the spectrum, $k_{\text{peak}} \Delta x/2\pi$, is displayed in 
Fig.~\ref{fig:nr-adjtime-kpeak}. To smooth out abrupt jumps of the peak
location, a ten-step moving average is taken for the latter.
In both cases, we adjust the starting time in the plot to
coincide with the instant when the bubbles first collide, allowing
for a meaningful comparison between runs with different parameters. 
Assuming that the bubble is expanding at roughly the speed of light, the 
starting time of bubble collision can be estimated as 
$t=\left(r_{s}-2r_{0}\right)/2$.

From Fig.~\ref{fig:nr-adjtime-kpeak}, we observe that
the peak of the spectrum is located around $k\Delta x/2\pi \approx 0.02$
independently of the initial bubble separation $r_s$ and of 
the lattice size $N$. 
This value corresponds to a wavelength of $\lambda_k \approx 50\Delta x$, 
or $m_H \lambda _k \approx 9$. 
On the other hand, the mean wave number of the spectrum does not show a clear 
dependence on either the initial bubble separation $r_s$ or the lattice size 
$N$. Nevertheless, curves with larger $r_s$ in Fig.~\ref{fig:nr-adjtime}
show relatively larger fluctuations.  
This is because for a given lattice size, larger values of $r_s$ lead to a 
fewer number of bubbles. 
For instance, for $N=256$ and $r_s=128$ there are only 4 bubbles in the 
lattice, which might result in  
significant statistical fluctuations. 

\section{Random bubble nucleation\label{sec:rand-nucl}}
\begin{figure}
\begin{centering}
\includegraphics[width=0.45\textwidth]{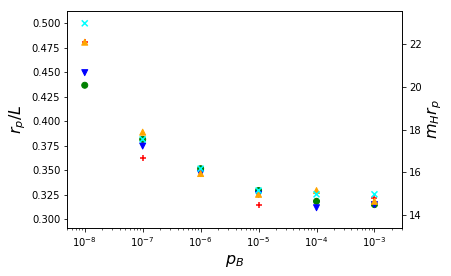}
\par\end{centering}
\caption{
Percolation size $r_p$ in units of
$L$ and in units of $1/m_H$ as a function of $p_B$. 
The percolation size is almost independent of $\gamma$. 
Different colors (and markers) in this plot indicate
different values of the Higgs damping $\gamma$:
red `+': $\gamma = 0$; 
green `o': $\gamma = 2.34\times 10^{-5}$;
blue `$\bigtriangledown$': $\gamma = 10^{-4}$;
orange `$\bigtriangleup$': $\gamma = 10^{-3}$;
cyan `x': $\gamma = 10^{-2}$.
}
\label{fig:rand-percolation-size}
\end{figure}

Having gained an intuition for the magnetic field resulting from multiple
bubble collisions, we are now in a position to model accurately a first-order
EWPT by allowing for bubbles to nucleate
at random locations in the unbroken phase.  
We perform 30 simulations on a lattice of size $N=256$, 
varying the nucleation probability over 6 
different values,
$p_{B}= 10^{-3}, 10^{-4}, 10^{-5}, 10^{-6}, 10^{-7}, 10^{-8}$,
and considering also 5 different values for the
Higgs damping, $\gamma=0, 2.34\times 10^{-5}, 10^{-4}, 10^{-3}, 10^{-2}$,
where the second one is equal to the experimentally measured decay width.
As can be seen from our results 
(Fig.~\ref{fig:rand-percolation-size}),
for $p_B \gtrsim 10^{-4}$
the total number of bubbles reaches the
maximum number of bubbles that the lattice can accommodate almost
instantaneously and 
the percolation size does not depend on $p_B$.  
Conversely, for $p_B \lesssim 1/2N^4 \approx 10^{-10}$, the nucleation probability is so small 
that only one bubble may be generated before the phase transition is completed in our lattice. In practice, already for $p_B=10^{-8}$ the results show a large variance because our lattice contains too few bubbles. 
To explore configurations with smaller $p_B$ would require a larger lattice.
In the rest of the paper, we shall focus on configurations with $10^{-7} \le p_B \le 10^{-4}$. 

To better understand the results of the simulations, 
we divide the evolution into three stages: 
\begin{enumerate}
\item[1.] {\it Free expansion (FE stage)}: the time before the nucleated bubbles collide 
	with each other. During this stage no magnetic field is generated.
\item[2.] {\it Bubble collision (BC stage)}: this stage starts when the bubbles collide 
	with each other, and the generation of magnetic field starts. 
	At this point, the broken phase does not yet extend to the whole 
		lattice.
\item[3.] {\it Higgs oscillation (HO stage)}: the Higgs field has 
	completely left the symmetric phase but is still oscillating, and 
		the generation of magnetic fields continues.
\end{enumerate}

Although we cannot draw clear lines between the three stages, the shift
from Stage 1 (FE stage) to Stage 2 (BC stage) is signaled by the onset of magnetic field generation. 
This typically occurs when $m_Ht\apprle 20$ in the simulations considered 
here, although, it can generically depend on the value of $p_B$. 
The boundary between Stage 2 (BC stage) and Stage 3 (HO stage) is roughly given by $T_{\text{stop}}$, 
which determines when the Higgs field at each lattice site has left the 
symmetric phase. 
This transition occurs when $m_Ht\sim200$ 
with some slight dependence on $p_B$ for the range of values that we consider.

We have carried out a set 
of ``bubble-collision stage'' simulations that focus on the magnetic
field generation during the phase transition.  
These simulations cover the first two stages, and the magnetic fields
generated up to $T_{\text{stop}}$ are analyzed. 
In addition, we have also performed 
``Higgs-oscillation stage'' simulations that are run until 
well into Stage 3, when the Higgs has completely left the symmetric phase
but is still oscillating and producing magnetic fields. We discuss them
in turn.

\subsection{Bubble-collision stage simulation}
\label{subsec:Short-time-simulation}
In this section the results of the 30 parameter combinations
defined at the beginning of Sec.\ref{sec:rand-nucl}
will be compared during Stages 1 and 2 of the phase
transition,  {\it i.e.} for $T\le T_{\text{stop}}$.
As mentioned above, the stopping point $T_{\text{stop}}$ of the simulations is 
chosen to be when $\vert\Phi\vert_{\text{MA10}}^{2}=0.25\eta^{2}$. 

\begin{figure}
\includegraphics[width=0.45\textwidth]{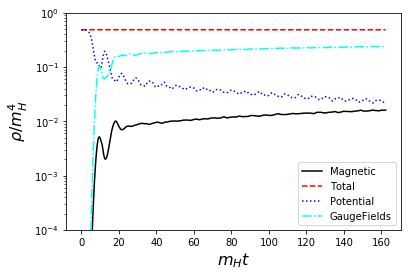}
\includegraphics[width=0.45\textwidth]{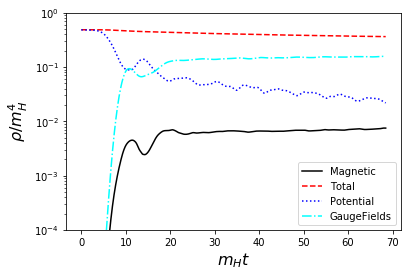}
\caption{
	Log-scale plots of magnetic energy, total energy, Higgs potential energy and energy in gauge sector (sum of $E_{U(1)}$ and $E_{SU(2)}$), for two different configurations:
({\it Top}): $\gamma=2.34\times10^{-5}$, $p_B=10^{-4}$; 
({\it Bottom}): $\gamma = 0.01$, $p_B=10^{-6}$. 
}
\label{fig:rand-single-energy}
\end{figure}

\begin{figure}
\subfloat[\label{fig:evo-B-gamma}$\rho_{B}\left(t\right)$ for different value of $\gamma$,
here $p_{B}=10^{-4}$. The curves stop at $T_{\text{stop}}$. ]{\begin{centering}
\includegraphics[width=0.45\textwidth]{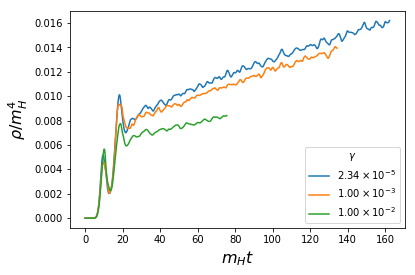}
\par\end{centering}
}

\subfloat[\label{fig:evo-B-p}$\rho_{B}\left(t\right)$ for different values of $p_{B}$, here
$\gamma=2.34\times 10^{-5}$. The curves stop at $T_{\text{stop}}$.]{\begin{centering}
\includegraphics[width=0.45\textwidth]{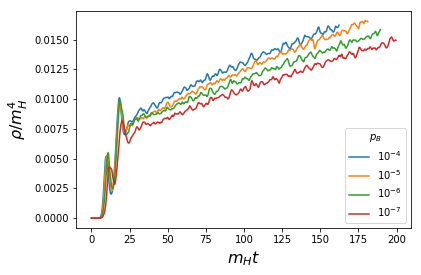}
\par\end{centering}
}

\caption{\label{fig:evo-B}
Plots of magnetic energy density as a function of $t$, $\rho_{B}(t)$, 
for different values of $p_B$ and $\gamma$, respectively. }
\end{figure}

In Fig.~\ref{fig:rand-single-energy}, various contributions to the
energy density, as well as the energy density of the generated 
magnetic field, are shown as a function of time. 
Two different parameter combinations, $\gamma = 2.34 \times 10^{-5}$
\& $p_B = 10^{-4}$ and $\gamma = 0.01$
\& $p_B = 10^{-6}$, are displayed. 
The total energy is not conserved due to the presence of the Higgs damping
term. Among the different contributions to the energy density, it is
the Higgs potential energy the one that shows a significant and sustained
decrease. This is in agreement with 
the expectation that the Higgs damping term is working properly
 only on the Higgs radial degree of freedom. The simulations
start with vanishing gauge fields. However, once bubbles are nucleated
and start to collide with each other, energy is transferred to the
gauge fields, and magnetic fields are generated as well.  
Although, in the period considered here, the energy in the magnetic field is 
only a few percent of the total energy, 
it does not stop increasing by the end of these runs as 
can be observed  
in Fig.~\ref{fig:evo-B}, where we plot the magnetic energy 
versus time for a selection of $p_B$ and $\gamma$ values. 
Fig.~\ref{fig:evo-B-gamma} shows that 
increasing the damping  
reduces the magnetic field.
This is so because the condition 
$\vert\Phi\vert_{\text{MA10}}^{2}\ge0.25\eta^{2}$ 
is met earlier due to the faster attenuation of the Higgs field, and
the duration of stages 1 and 2 is decreased. 
Also, for larger damping, a larger proportion of the total energy 
gets dissipated, thus reducing the energy available for magnetic field 
generation. 
This is clearly seen for $\gamma \apprge 10^{-4}$; while 
for $\gamma \apprle 10^{-4}$, we found the effects of
Higgs damping become negligible.
Fig.~\ref{fig:evo-B-p} shows that increasing $p_{B}$ increases both the 
magnitude and the rate $d\rho_B(t)/dt$ of magnetic energy density generated. 
Furthermore, we notice that for smaller $p_B$, 
the onset of magnetic field generation (i.e. the beginning of bubble 
collision) occurs later.
For $p_B \apprge 10^{-4}$, 
bubble nucleation, as well as magnetic field generation, is saturated on 
the lattice, and the corresponding curves that fall in this range
are similar to each other.

\begin{figure}
\begin{centering}
\includegraphics[width=0.45\textwidth]{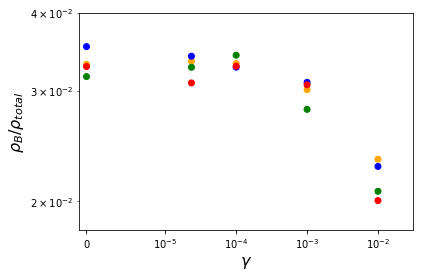}
\par\end{centering}
\caption{\label{fig:MagE.ratio.vs.damping}Magnetic energy conversion ratio
$\rho_{B}/\rho_{\text{total}}$ as a function of $\gamma$ at $T=T_{\text{stop}}$. 
Different colors correspond to different $p_{B}$ ($10^{-7} \le p_B \le 10^{-4}$).
}
\end{figure}

The dependence of the magnetic energy with the Higgs damping $\gamma$ 
is shown in Fig.~\ref{fig:MagE.ratio.vs.damping}
for different values of $p_{B}$.
When the damping falls below
$\gamma \apprle 10^{-4}$, magnetic field production reaches ``saturation''
and there is no dependence on $\gamma$. Indeed, choosing the damping equal
to the SM Higgs width makes almost no difference compared to the case with
no damping at all. 
 
Also from Fig. \ref{fig:MagE.ratio.vs.damping}, we deduce that
the ratio $\rho_{B}/\rho_{\text{total}}$ of the magnetic energy 
to the total energy at the end of the BC
stage, i.e. at $T=T_{\rm stop}$, turns out to be $\approx 2\%$-$4\%$.
More specifically, for $\gamma \apprle 10^{-4}$, 
this ratio saturates to the value $\sim 3.3\%$, 
while for $\gamma \apprge 10^{-4}$ it decreases with increasing Higgs damping,
approaching $\sim 2.3\%$ when $\gamma = 0.01$. 
These results are summarized in Table~\ref{tab:EM-conv-rate}.

\begin{table}
\begin{centering}
\begin{tabular}{|c|c|c|}
\hline 
$\gamma$ & $\rho_B/m_H^4$ & $\rho_{B}/\rho_{\text{total}}$\tabularnewline
\hline 
\hline 
$0.00$ & $0.0160$ & $0.0331$\tabularnewline
\hline 
$2.34 \times 10^{-5}$ & $0.0162$ & $0.0335$\tabularnewline
\hline 
$1.00 \times 10^{-4}$ & $0.0160$ & $0.0332$\tabularnewline
\hline 
$1.00 \times 10^{-3}$ & $0.0139$ & $0.0301$\tabularnewline
\hline 
$1.00 \times 10^{-2}$ & $0.00842$ & $0.0233$\tabularnewline
\hline 
\end{tabular}
\par\end{centering}
\caption{\label{tab:EM-conv-rate} 
The magnetic energy density measured in units of $m_H^4$ and the conversion rate 
$\rho_B/\rho_{\text{total}}$ for $p_B=10^{-4}$.
}

\end{table}

\begin{figure}
\begin{centering}
\includegraphics[width=0.45\textwidth]{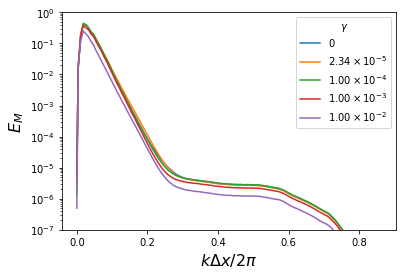}
\par\end{centering}
\caption{\label{fig:spec_vs_gamma}
Magnetic energy spectra for different values of $\gamma$ at $T=T_{\text{stop}}$. $p_B = 10^{-4}$. 
The range of $k\Delta x/2\pi$ is divided into 120 bins (thus $\Delta K'/N=0.0072$).
For $\gamma = 0$, the spectrum peaks at the position $k\Delta x/2\pi = 0.025$ (fourth bin);
for the other cases, the spectra peaks at the position $k\Delta x/2\pi = 0.018$ (third bin).
}
\end{figure}

\begin{figure}
\begin{centering}
\includegraphics[width=0.45\textwidth]{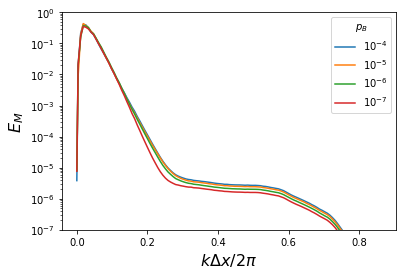}
\par\end{centering}
\caption{\label{fig:spec_vs_pb}
Magnetic energy spectra for different values of $p_B$ at $T=T_{\text{stop}}$. $\gamma = 2.34\times 10^{-5}$. 
For all the shown cases, the spectra peaks at the position $k\Delta x/2\pi = 0.018$ (third bin).}
\end{figure}

A more detailed picture of the generated magnetic field can be obtained
from Fig. \ref{fig:spec_vs_gamma}, which shows the field spectrum 
$E_M(k)$, defined in Eq. (\ref{eq:E_M}), for
various values of $\gamma$ at fixed $p_B = 10^{-4}$. 
The dependence on spectrum with $p_B$, for fixed $\gamma = 2.34\times 10^{-5}$,
is displayed in Fig. \ref{fig:spec_vs_pb}.
In both cases, the spectrum is shown as a function of the physical wave number 
$k\Delta x/2\pi = K'/N$, defined in Eq. (\ref{eq:wave-number}). 
The width of each bin is $\Delta k \Delta x/2\pi = \Delta K'/N$.  
The spectrum is normalized in such a way that the area under the curve is 
equal to the magnetic energy density. One caveat is that, the nonspherical
geometry of the lattice may lead to the underestimation of the spectrum
for $k\Delta x/2\pi \apprge 0.5$. 
The vanishing tail for large $k$ shows numerical artifacts from $k \sim 1/\Delta x$ are well controlled. 
It is clear from Figs. \ref{fig:spec_vs_gamma} and \ref{fig:spec_vs_pb}
that certain features of the spectrum are largely
independent of $p_B$ and $\gamma$.  
Specifically, the peak of the spectra lies at 
$k\Delta x/2\pi = 0.018$.
Thus, the dominant wavelength is 
$\lambda_k = 2\pi/k = \Delta x/0.018 \approx 56\Delta x$, 
or $m_H \lambda_k \approx 56 m_H \Delta x \approx 10$ 
($\lambda_k \approx 10^{-15}$ cm).

\subsection{Higgs-oscillation stage simulations
	\label{subsec:Long-time-simulation}}

At HO stage, the Higgs field has left the symmetric phase but it is
still oscillating around the minimum of the potential. As a result, 
magnetic fields can continue to be generated and this is what we set out 
to study in this section.

\begin{figure*}
\begin{centering}
\includegraphics[width=0.40\textwidth]{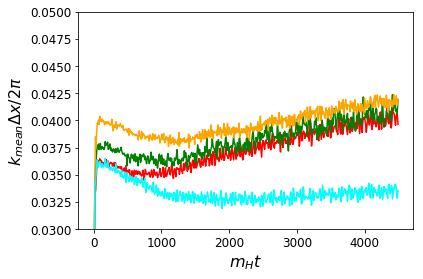}
\includegraphics[width=0.40\textwidth]{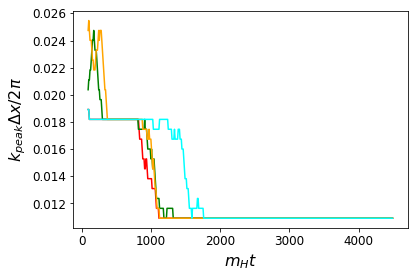}
\includegraphics[width=0.405\textwidth]{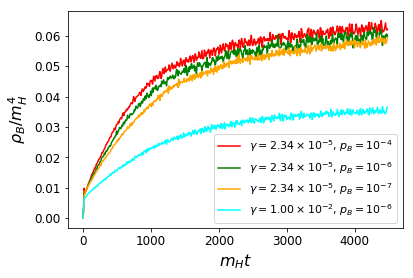}
\par\end{centering}
\caption{\label{fig:long-time-sim}
Plots of the results from ``Higgs-oscillation stage'' (HO stage) simulations. 
({\it Top left}): $k_{\text{mean}} \Delta x/2\pi$ as a function of time $m_Ht$;
({\it Top right}): $k_{\text{peak}} \Delta x/2\pi$ as a function of time $m_Ht$;
({\it Bottom}): Energy density of magnetic field, $\rho_B$, as a function of time $m_Ht$. The legends are the same for the three plots, and are only shown on the bottom plot.
}
\end{figure*}

To this effect we select a representative subset of the configurations 
considered before with the following set of parameters:
\begin{itemize}
\item[(i)] $\gamma=2.34\times 10^{-5}$, $p_{B}=10^{-4}$
\item[(ii)] $\gamma=2.34\times 10^{-5}$, $p_{B}=10^{-6}$
\item[(iii)] $\gamma=2.34\times 10^{-5}$, $p_{B}=10^{-7}$
\item[(iv)] $\gamma=1.00 \times 10^{-2}$, $p_{B}=10^{-6}$.
\end{itemize}
The evolution is followed for 100,000 time steps, $m_Ht\approx4500$, and
a snapshot of the configuration is saved every 200 time steps.
The outcome of this calculation is shown
in Fig.~\ref{fig:long-time-sim}. The upper-left plot displays 
$k_{\text{mean}}(t)\Delta x/2\pi$.
The upper-right plot shows $k_{\text{peak}}(t) \Delta x/2\pi$, the mode
of the spectrum, where, as before, the ten-step average is used.
Finally, the lower plot depicts the magnetic
energy density, $\rho_{B}(t)/m_H^4$.

First, we notice that the magnetic
energy density keeps increasing within the time range of the simulations. 
\begin{table}
\begin{centering}
\begin{tabular}{|c|c|c|c|}
\hline 
$\gamma$ & $p_B$ & $\rho_{B,BC}/m_H^4$ & $\rho_{B,HO^*}/m_H^4$ \tabularnewline
\hline 
\hline 
$2.34 \times 10^{-5}$ & $10^{-4}$ & $0.016$ & $0.062$\tabularnewline
\hline 
$2.34 \times 10^{-5}$ & $10^{-6}$ & $0.016$ & $0.060$\tabularnewline
\hline 
$2.34 \times 10^{-5}$ & $10^{-7}$ & $0.015$ & $0.060$\tabularnewline
\hline 
$1.00 \times 10^{-2}$ & $10^{-6}$ & $0.0075$ & 0.037\tabularnewline
\hline 
\end{tabular}
\par\end{centering}
\caption{\label{tab:final-em-density} Magnetic energy density at $t\sim T_{\text{stop}}$ and at the end of our simulation.
	} 
\end{table}
For example, for the configuration
with $p_{B}=10^{-4}$ and $\gamma=2.34\times 10^{-5}$ (red curve), 
before $m_Ht\sim1000$, the magnetic energy density $\rho_B$ 
grows roughly linearly with time;
although the rate slows down after that, 
$\rho_B$ keeps on increasing throughout the simulation time range. 
Keeping in mind that $m_H t_{\text{stop}}\sim150$ for this configuration,
it is clear that magnetic energy is indeed generated in the HO stage. 
In fact,  
the magnetic energy density at $m_Ht \sim 150$ is $\rho_B/m_H^4 \sim 0.016$, 
and it grows by a factor of $\sim 4$ to reach $\rho_B/m_H^4 \sim 0.062$ 
at the end of our simulation.
Hence, the generation of magnetic fields in the HO stage dominates over that in the BC stage.

The plot of $k_{\rm mean}$ and $k_{\rm peak}$ in Fig.~\ref{fig:long-time-sim} shows that the
magnetic energy has power on length scales that are much larger than the particle physics
scale $m_H^{-1} \approx 6\Delta x$.
For example, when $\gamma = 2.34\times10^{-5}$, independent of $p_B$, $k_{\text{mean}}\Delta x/2\pi$ 
converges to $\sim 0.04$, equivalent to a wavelength of $m_H\lambda_k \approx 4.2$. 
The power spectrum of the magnetic field peaks at even larger length scales. From the plot of
$k_{\rm peak}$ we see that the peak moves to larger length scales with time and at the
end of our run, $k_{\text{peak}} \Delta x/2\pi \approx 0.011$ for all parameters.
(The plot is jagged because of binning effects.) This corresponds to a wavelength of 
$ \lambda_k = 2\pi/k = \Delta x/0.011 \approx 91 \Delta x$ ($m_H \lambda_k = 15.2$). 

\begin{figure}
\begin{centering}
\includegraphics[width=0.45\textwidth]{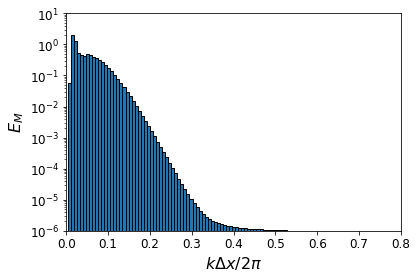}
\par\end{centering}
\caption{\label{pic:long-time-spec}
Spectrum of the magnetic field at $T=100000$ for a configuration with 
	$\gamma=2.34\times 10^{-5}$, $p_B=10^{-6}$. 
}
\end{figure}

In Fig.~\ref{pic:long-time-spec} we show the energy spectrum of the magnetic fields at the end of our simulation for $\gamma=2.34\times 10^{-5}$ and $p_B=10^{-6}$.
A peak is clearly seen in Fig.~\ref{pic:long-time-spec} and its location is largely independent of the parameters 
we varied in this paper. We conducted 
several runs on large lattices to test if the peak is due to finite lattice size and always found
the peak indicating the same wavelength, independent of the lattice size. 
Further study is needed to determine what parameters control the 
location and height of this peak.

\section{Discussion and Conclusion\label{sec:conclusion}}
We have simulated the classical dynamics of the bosonic electroweak theory to 
study the generation of magnetic fields during the EWPT, assuming 
that physics beyond the SM yields a first-order transition. 
Bubbles with the Higgs in the broken phase are randomly nucleated in regions 
where the Higgs is still in the symmetric phase, 
with the nucleation rate controlled by a parameter $p_B$ that we vary over a 
range. 
To account for the energy damped into fermions due to Higgs decays that will
be present in the full theory we add a damping term to the scalar EOM that
acts on the magnitude, $|\Phi|$, in a gauge invariant way.

We found it useful to divide the phase transition into three stages: 
		free expansion stage (FE), bubble collision stage (BC), 
		and Higgs oscillation stage (HO).
During the FE stage broken phase bubbles are nucleated and expand in the 
symmetric phase, but bubble collisions have not started yet and no
magnetic fields are generated. Once bubbles start crossing,
the energy density in magnetic fields grows to $\sim 3 \%$,
or $B \sim 10^{23}$ gauss, 
for $\gamma$ equal to the observed Higgs decay width.
Assuming the fields are nonhelical, the field strength will evolve to 
$\sim 10^{-15}$ gauss at present times~\cite{Banerjee:2004df}.
At this point the spectrum of the magnetic field 
has a peak at $k\Delta x/2\pi = 0.018$, 
or equivalently, the peak wavelength is $m_H \lambda_k = 10$ (or $\lambda_k \approx 2$ cm in comoving scale). 
This is larger than the initial size of the nucleated bubbles ($m_H r_0=6.5$) 
but not by much. 
We do not see a clear dependence of the peak location on either $\gamma$ or $p_B$. 
These findings are consistent with the results obtained in similar simulations (e.g. \cite{DiazGil:2008tf, Stevens:2012zz}).

Similar behavior was found in~\cite{DiazGil:2008tf,DiazGil:2007dy},
where the generation of magnetic fields in low-scale
electroweak hybrid inflation was studied on the lattice. In this context, 
electroweak symmetry breaking also occurs via the nucleation and growth of
Higgs bubbles and the system eventually enters a regime where magnetic
fields with energy density $\rho_B / \rho_{\text{total}}\sim 0.01$ were found.
Furthermore, as in our scenario (see Fig. \ref{fig:evo-B}),
$\rho_B$ was also observed to grow linearly with time. Let us emphasize however, that we consider initial conditions 
appropriate for a first-order phase transition with random bubble nucleation.
Unlike the scenario in~\cite{DiazGil:2008tf,DiazGil:2007dy}, our magnetic
field is initially zero and is entirely dynamically generated. 

A detailed characterization of the magnetic helicity is left for future work.
Nevertheless, let us point out that our equations of motion do not include
an explicit $CP$-violating term so we expect the average helicity to vanish.
Nevertheless, as observed in~\cite{DiazGil:2008tf,DiazGil:2007dy}, the
dispersion is expected to be nonzero leading to nonvanishing helical 
magnetic susceptibility.

After the BC stage, the evolution enters the HO stage in which 
the Higgs oscillates around its true vacuum for a very long time.
Magnetic energy is seen to continuously increase during the HO stage, even at the end of our simulation runs. Due to limitations in computation power, we are
not able to see the asymptotic value of magnetic energy density from our 
simulations. However, it is clear that most of the final energy in the 
magnetic field is produced during the HO stage, exceeding that produced in the
BC stage by a factor of $\sim 4$ (for typical values of $\gamma$ and
$p_B$) and still growing at the end of our runs.  
When decomposing the magnetic energy in Fourier space, we find the spectrum
to peak at
$k_{\text{peak}} \Delta x/2\pi = 0.011$, or $\lambda_k = 91 \Delta x$, 
comparable to the bubble size at percolation which, is approximately 0.33 of 
the lattice size (see Fig.~\ref{fig:rand-percolation-size}).
Our simulations suggest that the 
peak location is not sensitive to the damping
$\gamma$ nor the bubble nucleation probability $p_B$. This is consistent with
Fig.~\ref{fig:rand-percolation-size} in which we see that the percolation 
size is not sensitive to these parameters, in the range that we have
considered.

To summarize, using numerical simulations we find that a first-order EWPT
generates a significant amount of magnetic fields. 
While magnetic field generation has not stopped 
by the end of our simulations, we find that $\sim 10\%$ of the electroweak 
false vacuum energy is converted into magnetic fields. The energy spectrum of 
the magnetic field has a peak that shifts towards larger length scales. 
By the end of the BC stage, the peak wavelength is of the order of the bubble percolation scale, and 
it shifts to a longer wavelength in the HO stage. 

\acknowledgments

F.F. thanks the Institut de F\'{\i}sica d'Altes Energies and the 
Universitat Aut\`onoma de Barcelona for their
hospitality. T.V. thanks the Institute for Advanced Study for their 
hospitality.
F.F. and Y.Z. were supported in part by the U.S. Department of Energy, 
Office of High Energy Physics, under Award Number No. DE-SC0017987 at 
Washington University. 
T.V. is supported by the U.S. Department of Energy, Office of High Energy 
Physics, under Award No.~DE-SC0018330 at Arizona State University.

\bibliography{bubble_paper}

\end{document}